\documentclass[manuscript]{aastex}

\slugcomment{To appear in Astronomical J., 45.}

\shorttitle{Formation scenario} \shortauthors{Ortega et al.}

\begin{document}


\title{A formation scenario of young stellar groups in the region of the Scorpio Centaurus OB association}


\author{V. G. Ortega\altaffilmark{1}, E. Jilinski \altaffilmark{1, 2}, R. de la Reza \altaffilmark{1}, B. Bazzanella \altaffilmark{1}}


\altaffiltext{1}{Observat\'orio Nacional, Rua General Jos\'e
Cristino 77, S\~{a}o Cristov\~{a}o, 20921-400, Rio de Janeiro,
Brazil.}

\altaffiltext{2}{Pulkovo Observatory, Russian Academy of Science,
65, Pulkovo, 196140 St. Petersburg, Russia.}

\email{vladimir@on.br}

\begin{abstract}
The main objective of this work is to investigate the role played by
Lower Centaurus Crux (LCC) and Upper Centaurus Lupus (UCL), both
subcomponents of the Scorpio Centaurus OB association (Sco$-$Cen),
in the formation of the groups $\beta$ Pictoris, TW Hydrae and the
$\eta$ Chamaeleontis cluster. The dynamical evolution of all the
stellar groups involved and of the bubbles and shells blown by LCC
and UCL are calculated and followed from the past to the present.
This leads to a formation scenario in which (1) the groups $\beta$
Pictoris, TW Hydrae were formed in the wake of the shells created by
LCC and UCL, (2) the young cluster $\eta$ Chamaeleontis was born as
a consequence of the collision of the shells of LCC and UCL, and (3)
the formation of Upper Scorpius (US), the other main subcomponent of
the Sco$-$Cen association, may have been started by the same process
that created $\eta$ Chamaeleontis.

\end{abstract}


\section{Introduction}

The Scorpio Centaurus (Sco$-$Cen) OB association is one of the most
important sites of recent star formation in the solar neighborhood.
This association consists of three subgroups: Lower Centaurus Crux
(LCC), Upper Centaurus Lupus (UCL) and Upper Scorpius (US) (Blaauw
1964, de Zeeuw et al. 1999). According to their nuclear ages,
determined considering both high and low mass stars, LCC and UCL are
older with ages between 16 and 18 Myr (Sartori et al. 2003, Mamajek
et al. 2002) while US is younger with an age of about 5 Myr
(Preibisch and Zinnecker 1999, Preibisch and Mamajek 2008). It
should be noted that Blaauw (1978, 1991), using an independent
method of measuring ages based on stellar kinematics, found a
similar result for US. This already characterizes US as an unbound
stellar system.

The energetic output of massive stars in OB associations can
significantly affect the interstellar medium (ISM).An interesting
possible outcome of this interaction is the so-called triggered or
assisted star formation as opposed to spontaneous. Not only isolated
stars but entire groups of stars are thought to be formed by way of
the triggered star formation mode. The creation of expanding bubbles
and shells in the ISM by winds from massive hot stars and supernova
(SN) explosions are key ingredients in triggered star formation. In
fact, during the expansion the shell may become gravitationally
unstable due to increased density resulting from the accumulation of
swept-up gas and dust (Elmegreen and Lada 1977) or from the action
of SN explosions in the associated bubble (McCray and Kafatos 1987).
The instability fragments the shell leading to the formation of
dense molecular clouds and eventually to the formation of star
groups and clusters. Another possibility is that, during the
expansion, shock fronts associated with shells may compress clouds
existing in the ISM and ignite star formation. One interesting
characteristic of triggered star formation is that it can propagate
and form several star groups or associations. Increasing
observational evidences of this mode of star formation in our Galaxy
and in external nearby galaxies have been reported. Some
investigations on this subject include: Thronson et al. (1985),
Comeron et al.(1998), Comeron (2001), Oey et al. (2005), Deharveng
et al. (2005), Zavagno et al. (2006), Lee and Chen (2005), Lee and
Chen (2006), Chen et al. (2007), Carlson et al. (2007). Specifically
for the Sco$-$Cen association, the interaction of the stars with the
ISM and triggered star formation have been addressed by several
authors: Weaver (1979), Cappa de Nicolau and P\"{o}ppel (1986), de
Geus (1992). According to Preibisch and Zinnecker (1999) (see also
Preibisch and Mamajek 2008), the narrow range of ages observed in US
supports the view that the formation of this subgroup took place via
the triggering mode. An SN exploding in UCL is assumed to be the
triggering agent initiating the burst of star formation.

Several young stellar associations mainly composed of pre-main
sequence stars like $\beta$ Pictoris (BPMG) (Barrado y Navascu\'es
et al. 1999, Zuckerman et al. 2001, Ortega et al. 2002, 2004), TW
Hydrae (TWA) (de la Reza et al. 1989, Kastner et al. 1997, Webb et
al. 1999, Reid 2003, Mamajek 2005, de la Reza et al. 2006) and the
compact young cluster $\eta$ Chamaeleontis ($\eta$ Cha) (Mamajek et
al. 1999, 2000, Jilinski et al. 2005) are closely related to the
Sco$-$Cen OB association. The purpose of this work is to investigate
this relation with the aim to explore the part taken by LCC and UCL
in the formation of the BPMG, TWA, $\eta$ Cha and also US. In
principle this can shed light on the question of the way, or ways,
these stellar systems were formed. The approach we adopt is based on
the calculation of the past three-dimensional orbits of all the
systems involved. We also compute the evolution of the shells
associated with the bubbles created by the subcomponents LCC and
UCL. The expectation is that a formation scenario of the groups
BPMG, TWA and of $\eta$ Cha can be obtained by following their
temporal evolution combined with the evolution of LCC, UCL and the
shells of these Sco$-$Cen subcomponents. We expect that this
scenario will help us answer to several general questions concerning
this problem: what was the formation mode of the loose unbound
associations the BPMG, TWA and of the compact cluster $\eta$ Cha?
Were these systems originated in the shells formed by the bubbles
blown by LCC and UCL? Or did the birth of these stellar groups occur
exterior to the shells? Finally, had the formation of US something
to do with this process? A related point concerns the role played by
the SNe that certainly existed in LCC and UCL. Did the formation of
the groups take place as a consequence of the direct action of SNe
on the shells? Or was this action more indirect by contributing to
the formation of the bubbles and shells? All these questions are
important and, obviously, one cannot hope to obtain detailed answers
to all of them by only using stellar and shell dynamics. Nonetheless
this approach can provide a valuable general picture of star
formation in the region of the Sco$-$Cen association.

\section{The method}

The methodology of calculating the stellar three-dimensional past
evolution of young associations has been employed by us in previous
works (Ortega et al. 2002, 2004, Jilinski et al. 2005 and de la Reza
et al. 2006) to get estimates of their ages and places of origin.
This is realized by backward integration of the orbits of all the
stars of the groups taking into account the (modeled) gravitational
potential of the Galaxy. The region of first maximum concentration
of the orbits (confinement) we consider to be the birthplace of the
stellar group; the time interval, from today, we consider to be the
age of the group. A more detailed analysis of the orbit confinement
region leads to a star distribution pattern at birth that can be
considered as a representation of the density distribution in the
natal cloud. One example is the BPMG, whose dynamical age of 11.2
Myr can serve as calibrator for other pre-main sequence systems
which lack age determination but have similar youth features. This
can be of importance, for instance, in the investigation of the
temporal evolution of associated protoplanetary disks. Apart from
getting age estimates, this method is very efficient in detecting
intruding stars. This method can also be used to look for new,
potential, members of a group on the basis of the confinement of
their past orbits in the previously determined formation region of
the group. Furthermore, the confinement must take place at the
previously determined age of the group under investigation. This
condition is far more stringent than the simple comparison of the
present-day kinematics.As an example, using this technique and a
compilation, brought together by us containing more then 30,000
stars with Hipparcos entries and radial velocity measurements, we
were able to pick five new potential members of the BPMG. They are
included in Table 1 where we briefly comment on them. To confirm, or
reject, the membership of these systems to the BPMG, further
properties, typical of the group, should be investigated. In Table 2
we list the stars which likely are related to the BPMG. Although
they do not find themselves in the confinement region of the BPMG,
they are spatially in its neighborhood at the age of BPMG. They may
have formed not in group but in a more isolated fashion. To these
systems we ascribe the same age of the BPMG. Table 3 contains the
stars which very probably have no relation to the BPMG because their
orbits take them far away from this group. They can be considered as
interlopers.

A similar exercise realized for TWA with an age of 8.3 Myr did not
give any additional potential members, probably because of the small
number of stars with full kinematic data in the association.

The existence of two modes of star formation, in group and isolated,
seems to be quite common in associations, as in Orion complex(Lee
and Chen 2006) for example. Can we identify some formation mechanism
capable of giving origin to these modes? In the following, we shall
present a formation scenario in which such a mechanism could exist.

Stellar and shell kinematics have been used to study the velocity
distribution of stars in expanding shells (Moreno et al. 1999) and
also as a tracer of triggered star formation (Comeron et al. 1998,
Comeron 2001). In the next section we shall investigate the
evolution of stars and shells in the region of the Sco$-$Cen OB
association.A distinctive feature of our work is that we take into
account the temporal evolution of structures from the past to the
present.

\section{Dynamical evolution of the star groups and shells}

In Figues 1, 2, 3 we show the past positions of the stellar groups
the BPMG, TWA and $\eta$ Cha at the epochs of their formation as
obtained by us in previous works (Ortega et al. 2002, 2004, Jilinski
et al. 2005, de la Reza et al. 2006) using the method mentioned in
the previous section. In these figures the axes $X, Y, Z$ are
positive oriented in the directions of the Galactic center, the
Galactic rotation and above the Galactic plane respectively. The
positions of the stellar subcomponents LCC and UCL are also shown in
these figures. A common and intriguing aspect of these plots is the
location, at birth, of the BPMG and TWA behind LCC and UCL while
$\eta$ Cha shows a somewhat different relative configuration. What
is the origin of such groups disposition? Is it possible to find a
formation scenario compatible with such a disposition? (Ortega et
al. 2006)To investigate this question we consider the evolution of
the bubbles and shells originated by LCC and UCL in addition to
their stellar components.

The bubbles are blown by the combined action of the winds of hot
stars and SNe. To compute the effective resulting "mechanical
luminosity" we use the contribution of stellar winds and the number
of expected SNe in LCC and UCL found by de Geus (1992) and take 18
Myr as a mean age for both subcomponents. In the case of an
association it is necessary to take into consideration the fact that
the creation of a common, for all the stars, bubble is not a point,
instantaneous process, but one extended in space and time. Every
star of the association, mainly the hot ones, is contributing to
this process. A common, for the whole association, shock can be said
to have formed when the stars of the association turn out to be
within it. This fixes the initial radius $R_i$ and the initial time
$t_i$ for the expansion of the shock. The mass of the ambient medium
will be swept up by the common shock thereafter for $t > t_i$ and we
assume that this takes place in a uniform ISM of number density 100
 cm$^{-3}$ and molecular weight 2.8. To follow the time evolution
of the shells created by the UCL and LCC we integrate numerically
the system of equations given by Castor et al. (1975):

$$\frac {d}{dt}(\frac {4}{3} \cdot \pi \cdot R^3 \cdot \rho_0 \cdot
\frac {dR}{dt}) = 4 \cdot \pi \cdot R^2 \cdot P$$

$$\frac {dE}{dt} = \mathfrak{L} - P \cdot \frac{d}{dt}(\frac {4}{3}
\cdot \pi \cdot R^3) $$

$$E = 2 \cdot \pi \cdot R^3 \cdot  P$$

where $R$ is the radius, $P$ is the pressure exerted on the shell by
the hot bubble of thermal energy $E$, $\rho_0$ is the ISM density,
and $\mathfrak{L}$ is the "mechanical luminosity", that is, the rate
at which the energy is generated in the bubble.

Values for the initial thermal and kinetic energies were calculated
from the solution of Weaver at al. (1977) (see also Mac Low and
McCray 1988). Making use of the initial mass of the shell computed
as $M_i = 4/3 \cdot \pi \cdot R_i^3 \rho_0$, we determine a value
for the initial velocity of the shell. Solutions for the expansion
of the shells can be computed with and without the contribution of
the SNs to the mechanical luminosity. In both cases, we take $R_i =
15$ pc as the initial radius and start the integration at the
initial time $t_i = 5$ Myr.

\section{The formation of the stellar groups}

Figure 4(a) shows the solutions for the shells of UCL and LCC
without the contributions of the SNs. From this figure we see that
although there is a certain approach between the shells, they do not
come into contact.

Table 4 lists and Figure 4(b) shows the details of the solution (the
radii and velocities of the UCL and LCC shells) for the case where
both, stellar winds and SNs, contribute to the mechanical energy
supplied to the bubbles. As in the previous case the shells
gradually approach each other but now, about 9 Myr, they have met
together starting the process of interaction between them (Figure
7). In the narrowing process of the approaching shells a supersonic
flow is expected to arise in the funnel created by them. We identify
the flow so formed as a physical mechanism which could be
responsible for the formation of the BPMG and TWA (Figures 5 and 6).
In the case of TWA an additional trigger can be present besides the
pressure field originated by the funneling of the shells: in fact, a
Mach shock is expected to arise as a result of the shells collision
(Figure 7). Such mechanisms can explain quite naturally why the
stellar groups the BPMG and TWA were born in the wake of LCC and
UCL. They can also explain the formation of isolated star systems,
as is the case in the neighborhood of the BPMG.

Shells collisions have been proposed by Chernin et al. (1995) as a
mechanism of violent star formation. Characteristic of this process
is the appearance of two reflected shocks which, dragging material
from the region of collision, move to the internal areas of the
bubbles giving rise to the formation of "champagne flows". In Figure
8 we show the position of the $\eta$ Cha cluster at the time of its
birth, 6.7 Myr ago, determined previously by us (Jilinski et al.
2005); $\eta$ Cha is at the shell of LCC and by this time the
"champagne flow" has advanced into the LCC bubble. We identify the
"champagne flow" in this region as the mechanism triggering the
formation of the $\eta$ Cha cluster.

Another interesting point refers to the reflected shock which surged
into the bubble of UCL in a direction symmetric relative to the
first one. In Figure 9 we show the situation at $-5$ Myr, the age of
US formation (Preibisch and Zinnecker 1999, Preibisch and Mamajek
2008). At this time point US is in the shell of UCL. The
configuration is such that the "champagne flow" created by the
reflected shock should have interacted with the shell of UCL
igniting the burst of star formation which may have given origin to
the US subcomponent.

\section{Discussion and conclusions}

In the Introduction we posed some questions concerning the origin of
young stellar groups in the region of the Sco$-$Cen OB association
and wondered whether using stellar and shell dynamics, considered in
their time evolution, would give us clues to tackle those questions.
We found that this approach leads to a picture in which quite
different physical mechanisms capable of inducing star formation can
occur. According to this picture, the unbound groups the BPMG and
TWA were born in regions of the medium between LCC and UCL, the
source of overpressure being the flow generated by the approaching
shells of these subcomponents and the Mach shock arisen after the
shells have come together. On the other hand, the formation of the
compact cluster $\eta$ Cha took place quite differently, the trigger
being one of the energetic "champagne flows" arising as a result of
the collision of the shells of LCC and UCL. Are such formative
differences reflected in the observed properties of the BPMG, TWA,
and $\eta$ Cha? This seems in fact to be the case.

Table 5 contains, and Figures 10, 11, and 12 show the velocity
components of the stellar groups the BPMG, TWA, and $\eta $ Cha
relative to the local standard of rest (LSR) at the epochs of their
formation. The velocity components of LCC and UCL are also included.
All the groups have $V_y$ components negative, that is, contrary to
the Galactic rotation. This reflects the peculiar motion of the gas
complex. At the same time, all the groups have $V_x$ components
positive (the direction to the Galactic center), that is, in the
course of time they will loose rotational support until the Galactic
rotation takes over. How does the situation at these epochs look
relative to the average motion of LCC and UCL? Table 6 and Figure 13
present the velocity components of the BPMG, TWA and $\eta$ Cha in
this reference system at the times of their formation. We see that
the BPMG has the largest positive $V_y$ component, equal to 2.4 km
s$^{-1}$. This is consistent with the proposed mechanism of gas flux
produced by compression and also explains why the BPMG is farther
away from LCC. On the other hand, the $V_y$ component of TWA is also
positive but significantly smaller than that of the BPMG which is
consistent with the final stage of compression resulting from the
shells collision 9 Myr ago. Finally, the $V_y$ component of $\eta$
Cha is negative and quite small, whereas its $V_x$ component is
larger and, in the Galactic anticenter direction, consistent with a
triggering action due to the "champagne flow". The sequence shown by
Figure 13 strongly suggests the occurrence of one process, involving
several triggering mechanisms, taking place during the temporal
evolution of the Sco$-$Sen subcomponents LCC and UCL.

In Figure 14 we show the situation at the epoch 13 Myr ago, when the
common shell was formed and the region where the BPMG would be
formed at about 11 Myr ago. It can be wondered whether such 2 Myr
time intervals would be sufficient to produce the necessary
compression of the medium between LCC and UCL to induce the
formation of the BPMG. Couldn't a different, additional more
powerful triggering mechanism, be involved in the formation of that
stellar group? A SN event occurring in LCC prior to the formation of
the common shell, for example? Note that this would also be
consistent with the dynamical constraints and would not contradict
Figure 13. Such a possibility was investigated by us in a previous
work (Ortega et al. 2004) where an attempt was made to identify the
probable SN responsible for the formation of the BPMG.
Unfortunately, large uncertainties in the radial velocity of the
suspected runaway star made this attempt inconclusive. Better radial
velocities are needed in order to get reliable three-dimensional
orbits of runaway stars.

Aside from the bound nature of $\eta$ Cha, further differences have
been reported in the literature. Moraux et al. (2007), for example,
emphasize its very compact configuration, the absence of wide
binaries and mass segregation in this young cluster. All these
features point to a violent formation of $\eta$ Cha. Interestingly,
the position of the US subcomponent in the shell of UCL 5 Myr ago
follows from our past orbits calculations of US using the present
positions and velocities of its stars, and also from the modeled
evolutions of the shells.  It is worth noting that the formation of
US in the shell of UCL may have been triggered by the ``champagne
flow'' arising from the collision of the shells of LCC and UCL.

It should be stressed that the ambient mean number density of 100
cm$^{-3}$ used in the calculations is not arbitrary because it must
satisfy dynamical constraints set by the orbits of the stellar
groups and the evolution of the shells.

How does the mass swept up by the shells with the mass observed
today in the region of the Sco$-$Cen association? The masses of the
shells up to -9.0 Myr, the time when they collided, given by our
solutions are $6.4 \times 10^5 M_{\odot}$ for LCC and $8.0 \times
10^5 M_{\odot}$ for UCL (see Table 4). This gives a total mass of
$1.4 \times 10^6 M_{\odot}$, consistent with the value of about
$10^6 M_{\odot}$ quoted by Weaver (1979). We compare this value with
the mass seen today in the adjacent regions of LCC and UCL. For the
H I loops surrounding today the subcomponents LCC, UCL and US, de
Geus (1992) found a mass of $4.8 \times 10^5 M_{\odot}$. In addition
to this there is mass in the form of molecular clouds. In Figure 15,
constructed using the catalog of Dutra and Bica (2002), we show the
present-day situation. One identifies molecular gas aggregates such
as the Chamaeleontis clouds, the Lupus clouds, the Ophiuchus
complex, the Coma Australis molecular cloud and others. According to
Table 1 of Ballesteros-Paredes and Hartmann (2007), the mass in
these molecular clouds amounts to $0.8 \times 10^5 M_{\odot}$, which
added to the previous $4.8 \times 10^5 M_{\odot}$ H I mass found by
de Geus (1992) gives a mass of $5.6 \times 10^5 M_{\odot}$. In
addition to this, the Aquila Rift molecular cloud, with a mass of
$2.7 \times 10^5 M_{\odot}$, has been often genetically related to
the Sco$-$Cen association (for example Strai\v zys et al. 2003).
Then the overall identifiable mass in the region will be $8.3 \times
10^5 M_{\odot}$. If we do not include in this value the mass $1
\times 10^5 M_{\odot}$ for the LCC  H I loop found by de Geus
(1992), the resulting value $7.3 \times 10^5 M_{\odot}$ is
consistent with the mass associated with the UCL shell. As regards
LCC, quite a lot of mass would remain unidentifiable today and the
scenario here proposed would be indicating that a sizeable quantity
of mass has been leaving the system during the last 5-6 Myr, most
probably in the direction of the Galactic anticenter (as suggested
by Figures 8 and 9), a low density region in which the Sun is
located at present. In this respect, it is pertinent to note that de
Geus (1992) in his analysis of the present-day situation found no
evidence of expanding gas associated with the LCC subcomponent.

\acknowledgments

E.G.J thanks FAPERJ and MCT/Brazil for the financial support under
the contracts E-26/153.045/2006 and 384222/2006-4.

\begin{table}
\caption{Dynamical members of BPMG\label{tbl-2}}
\begin{tabular}{rrrrrrr}
\tableline\tableline HIP 560 & HIP 21547 &  HIP 27321  & HIP 92024 & HIP 92680 \\
HIP 103311 & HIP 10680 & HIP 11437A & HIP 84586 & HIP 12545 \\
HIP 14361\tablenotemark{a} & HIP 23200\tablenotemark{b}&
HIP 23309\tablenotemark{c}& HIP 25486 & HIP 99273\tablenotemark{d} \\
BD -17 6128 & HIP 105441\tablenotemark{e}& HIP 102409 & HIP 88399 & HIP 102141 \\
\tableline
\end{tabular}
\tablenotetext{}{This table contains stars whose 3D orbits confine
at the age of 11.2 Myr forming a group. The symbols a, b, c, d, e
refer to five new potential members.}

\tablenotetext{a}{Poorly known F5V star}

\tablenotetext{b}{Well known M type star V1005 Ori which
independently has been recognized as a member of BPMG by Torres et
al. (2006)}

\tablenotetext{c}{This star has been independently proposed by
Torres et al. (2006) as a member of BPMG}

\tablenotetext{d}{Also independently proposed by Moor et al. (2006)
as a member of BPMG}

\tablenotetext{e}{This star has been proposed as a possible member
of the Tucana association by Zuckerman et al. (2001). Here we
proposed this star as a member of BPMG}

\end{table}

\begin{table}
\caption{Stars related to BPMG}
\begin{tabular}{rrr}
\tableline\tableline HIP 23418AB & HIP 29964 & HIP 95270  \\
\tableline
\end{tabular}
\tablenotetext{}{The stars in this table do not confine but are
spatially related to BPMG at the age of 11.2 Myr.}
\end{table}

\begin{table}
\caption{Intruding stars not members of BPMG}
\begin{tabular}{rrrr}
\tableline\tableline
HIP 10679 & HIP 79881 & HIP 88726 & HIP 95261 \\
\tableline
\end{tabular}
\end{table}

\newpage

\begin{table}
\caption{Size and velocity evolution of the LCC and UCL shells}

\begin{tabular}{ccccccc}
\tableline\tableline

 &   LCC &   & & & UCL  &    \\
\tableline
Time & R$_{LCC}$ & V$_{LCC}$ & M$_{LCC}$ & R$_{UCL}$ & V$_{UCL}$ & M$_{UCL}$ \\
\tableline
 Myr &  pc & $ km \cdot sec^{-1}$ & M$_\odot \cdot 10^5$& pc & $km \cdot sec^{-1}$ & M$_\odot \cdot 10^5$ \\
\tableline

          -13 & 15.0  & 13.2  & 0.98 & 15.0 & 18.6  &  0.98 \\
          -12 & 22.0  &  4.2  & 1.36 & 23.6 &  4.8  &  1.67 \\
          -11 & 25.4  &  2.7  & 1.43 & 28.0 &  3.0  &  1.87 \\
          -10 & 27.8  &  2.1  & 1.36 & 30.1 &  2.3  &  1.77 \\
           -9 & 29.7  &  1.7  & 1.28 & 32.3 &  1.9  &  1.67 \\

\tableline
              & & $\sum M =$&  6.41 & & $\sum M =$ & 7.96 \\

\tablenotetext{}{Time scale presents time interval from today.}
\tablenotetext{}{R$_{LCC}$ and R$_{UCL}$ show the radii of the LCC
and UCL bubbles.} \tablenotetext{}{V$_{LCC}$ and V$_{UCL} - $show
the velocities of shell's expansions.} \tablenotetext{}{M$_{LCC}$
and M$_{UCL}$ show mass, initial plus swept-up in the shells
expancion.}

\tablenotetext{}{ Values of $\sum M $ present the total masses
accumulated during the shells expansion.}

\end{tabular}

\end{table}

\begin{table}
\begin{center}
\caption{Space velocity components (in km $\cdot$ s$^{-1}$) relative
to the Local Standard of Rest at the formation epochs of the BPMG at
-11.2 Myr, of TWA at -8.3 Myr and of the $\eta$ Cha cluster at -6.7
Myr}
\bigskip
\begin{tabular}{ccccccrcr}
\tableline\tableline
Name      & &   U   &  &  V   &  &   W   &  &   Age  \\
\tableline
         &  &  &   &  [km $\cdot$ s$^{-1}$]  &   &    &    &  [Myr]       \\
\tableline\tableline

LCC         &  &   2.8 &  & -11.9  &  &  1.4  &  &   -6.7  \\
UCL         &  &   4.6 &  & -13.1  &  &  2.1  &  &         \\
$\eta$ Cha  &  &   1.1 &  & -13.1  &  & -2.8  &  &         \\
\tableline
LCC         &  &   2.8 &  & -12.2  &  &  1.5  &  &   -8.3  \\
UCL         &  &   4.3 &  & -13.6  &  &  2.2  &  &         \\
TWA         &  &   1.2 &  & -12.4  &  &  1.0  &  &         \\
\tableline
LCC         &  &   2.9 &  & -13.1  &  &  1.6  &  &  -11.2  \\
UCL         &  &   4.1 &  & -14.8  &  &  2.4  &  &         \\
BPMG        &  &   0.3 &  & -11.5  &  & -2.5  &  &         \\

\tableline\tableline
\end{tabular}
\end{center}
\end{table}

\begin{table}
\begin{center}
\caption{Space velocity components (in km s$^{-1}$) relative to the
average motion of LCC and UCL at the formation epochs of the BPMG at
-11.2 Myr, the TWA at -8.3 Myr and of the $\eta$ Cha cluster at -6.7
Myr}
\bigskip
\begin{tabular}{ccccccrcr}
\tableline\tableline
Name      & &   U   &  &  V   &  &   W   &  &   Age  \\
\tableline
         &  &  &   &  [km s$^{-1}$]  &   &    &    &  [Myr]       \\
\tableline\tableline

$\eta$ Cha  &  &  -2.5 &  & -0.6   &  & -4.5  &  &   -6.7  \\
TWA         &  &  -2.4 &  &  0.5   &  & -0.8  &  &   -8.3  \\
BPMG        &  &  -3.2 &  &  2.4   &  & -4.6  &  &  -11.2  \\

\tableline\tableline
\end{tabular}
\end{center}
\end{table}

\begin{figure}
\epsscale{1.5} \plottwo{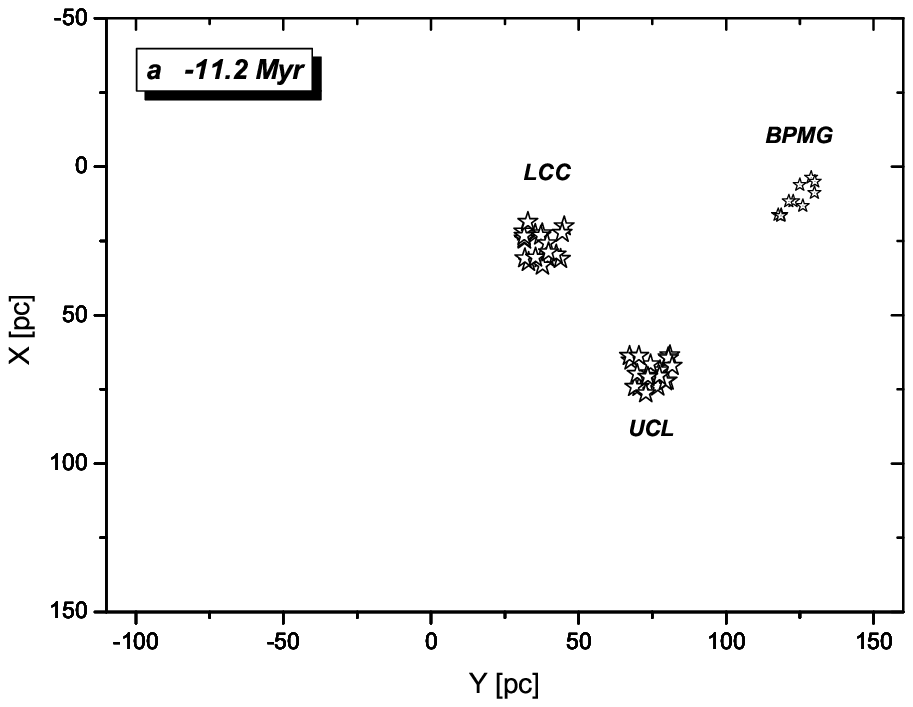}{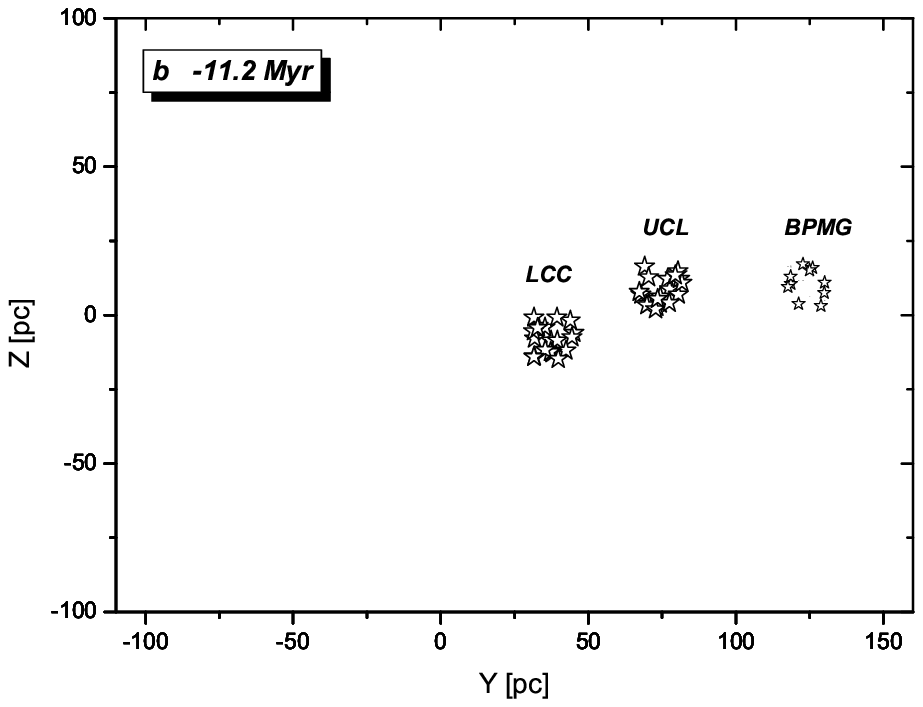} \caption{LSR positions of
LCC, UCL and BPMG group at the age of -11.2 Myr in the (X,Y) plane
(\textbf{a}) and in the (Y,Z) plane (\textbf{b}). The symbols
representing the stars are used to mark roughly the shape of each
group.\label{fig1}}
\end{figure}

\begin{figure}
\epsscale{1.5} \plottwo{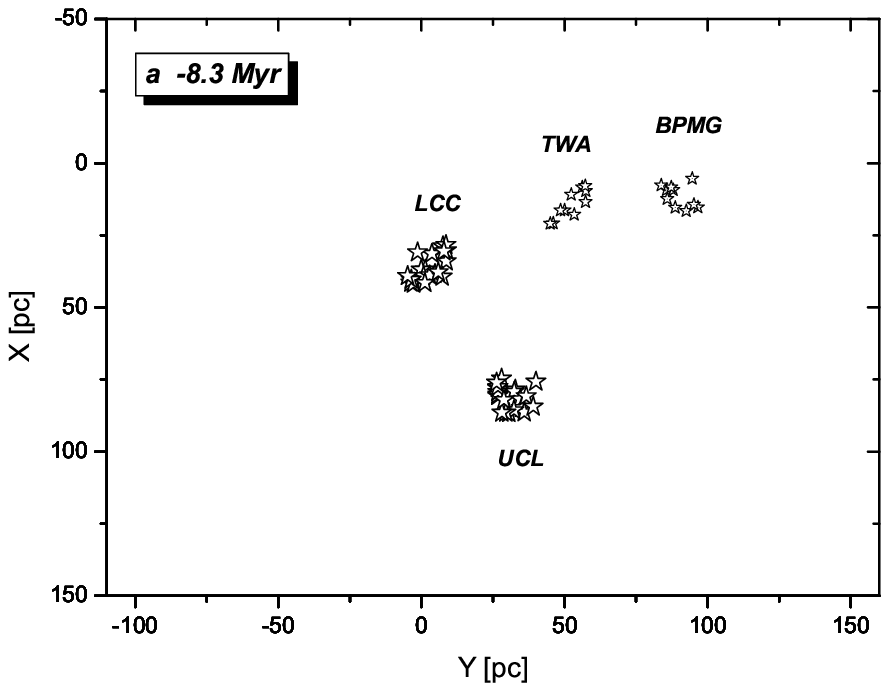}{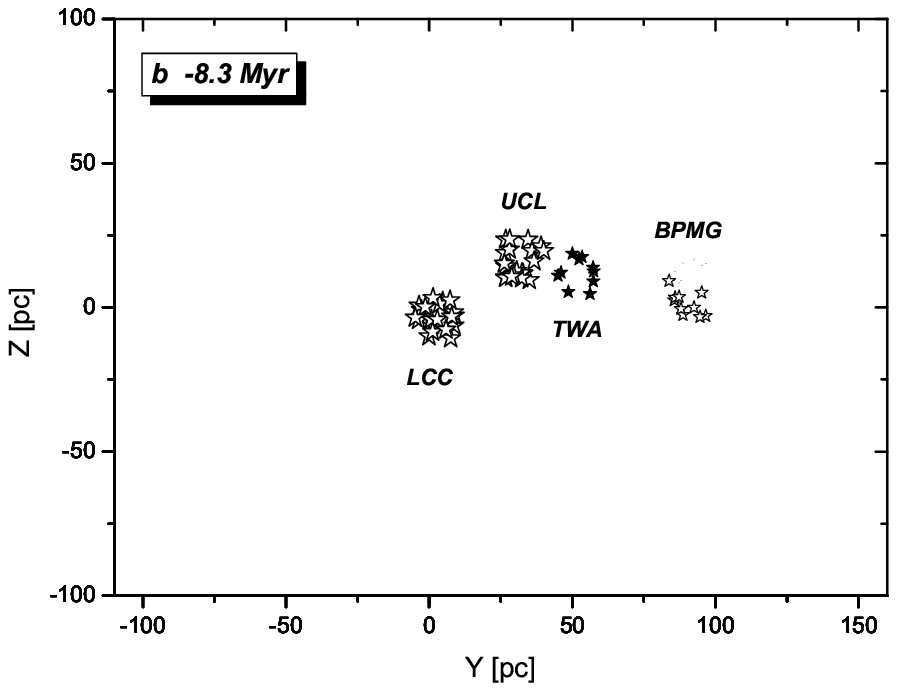} \caption{LSR positions of
LCC, UCL, BPMG group and TWA at the age of -8.3 Myr in the (X,Y)
plane (\textbf{a}) and in the (Y,Z) plane (\textbf{b}) The symbols
representing the stars are the same as in Fig. 1. \label{fig2}}
\end{figure}

\begin{figure}
\epsscale{1.5} \plottwo{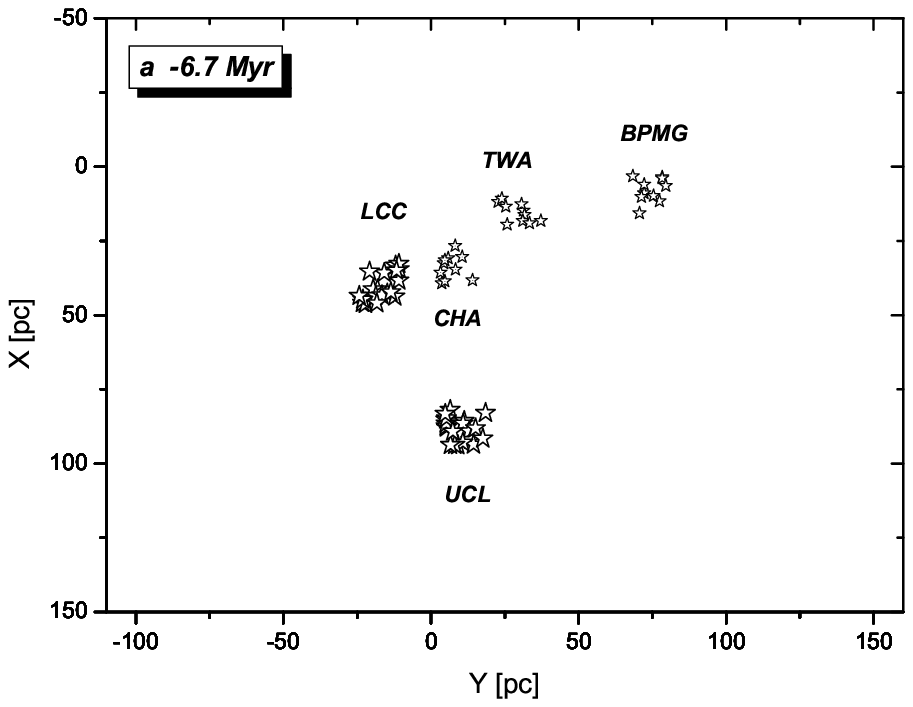}{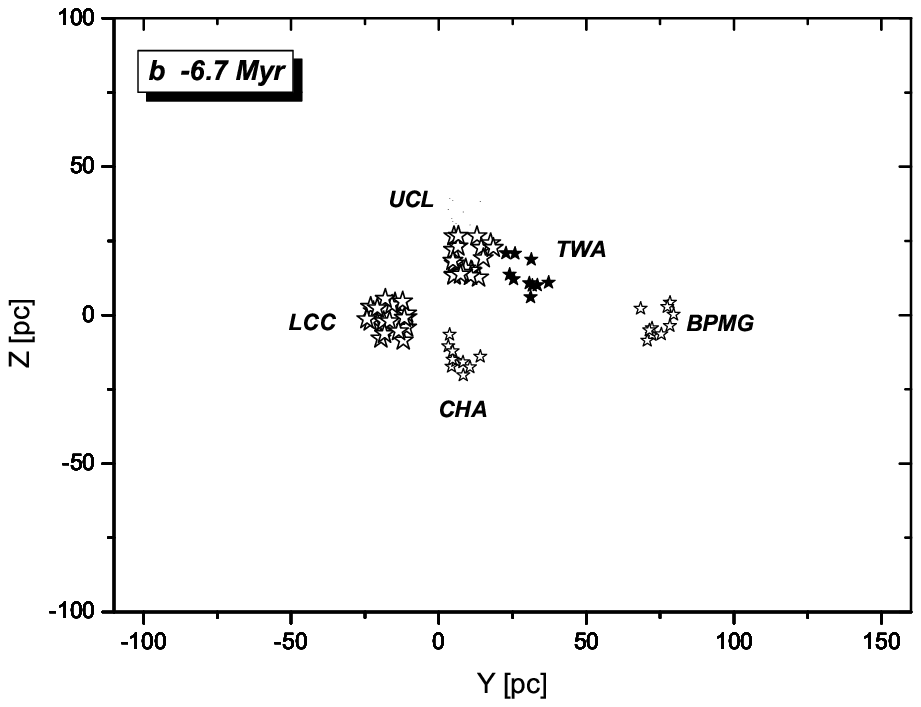} \caption{LSR positions of
LCC, UCL, BPMG group, TWA and $\eta$ Cha cluster at the age of -6.7
Myr in the (X,Y) plane (\textbf{a}) and in the (Y,Z) plane
(\textbf{b}).The symbols representing the stars are the same as in
Fig. 1. \label{fig3}}
\end{figure}

\begin{figure}
\epsscale{1.0} \plottwo{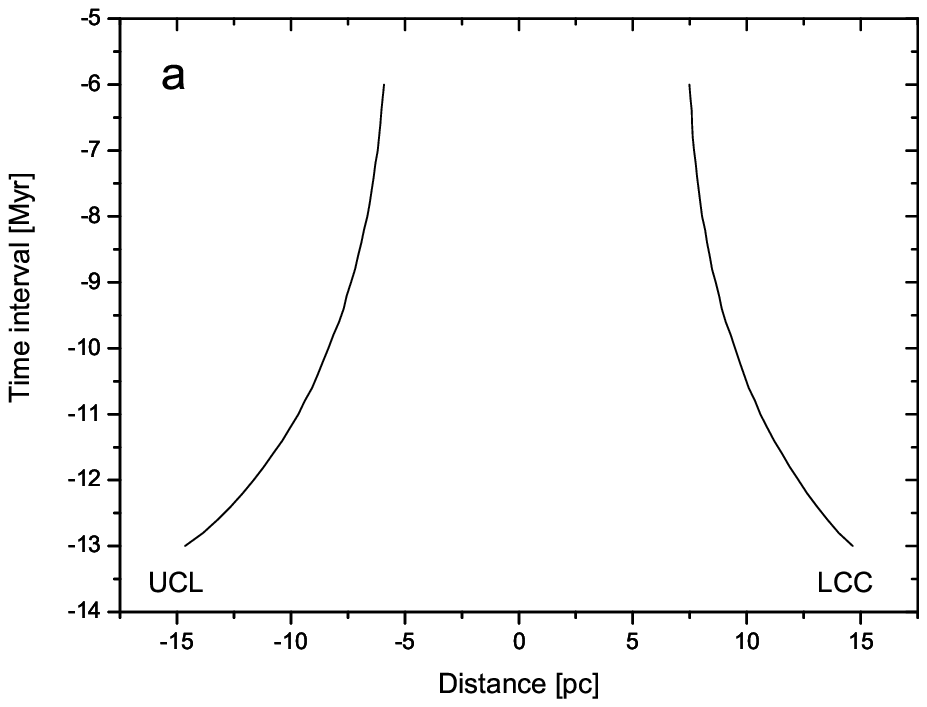}{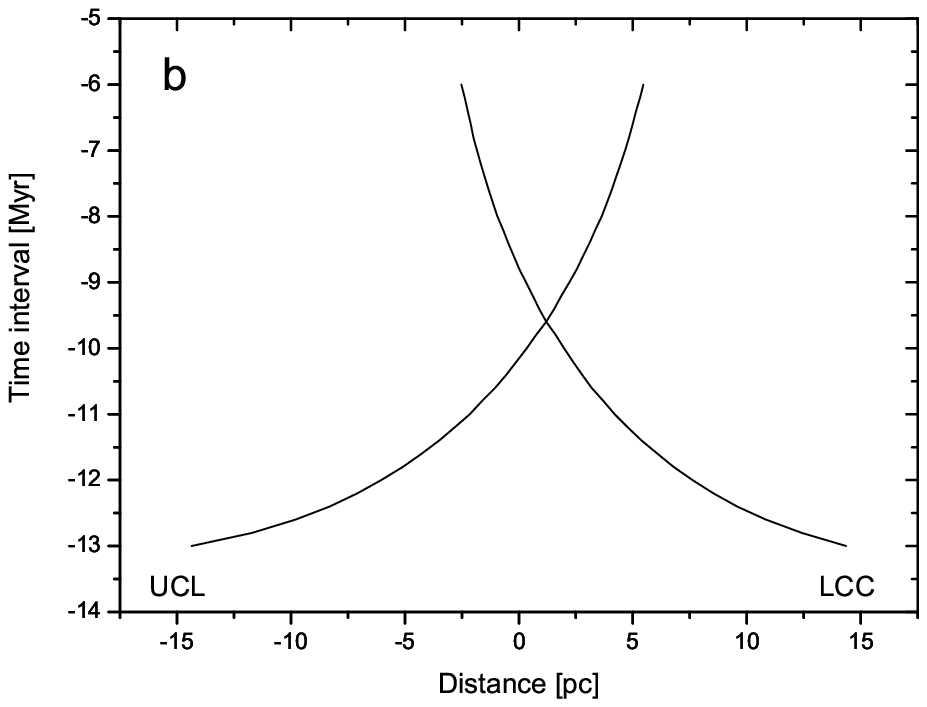} \caption{Time evolution
of the LCC and UCL shells. Left panel shows the shells evolution
without the SNs contributions. The right shows the shells evolutions
with the contribution of the stellar winds and SNs. The ordinate is
the time interval in Myr from today as in Table 4.The zero-point of
the distance scale corresponds to the mid-point between the stellar
centroids of UCL and LCC at each epoch. Each curve displays the time
evolution of the shortest distance from that mid-point to the
corresponding shell. The crossing point corresponds to the epoch of
the shells collision.\label{fig4}}
\end{figure}

\begin{figure}
\epsscale{.80} \plotone{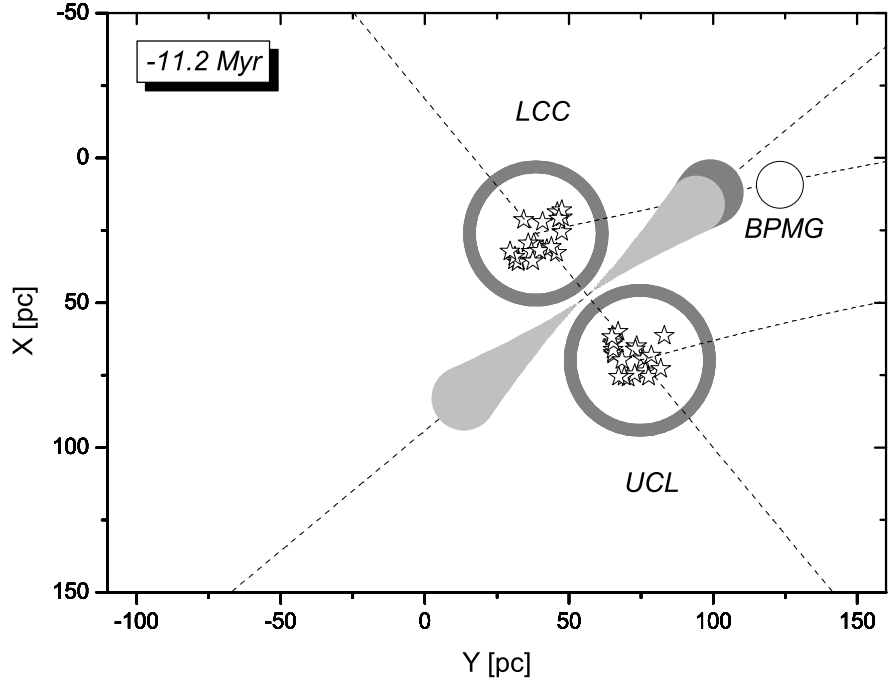} \caption{LCC and UCL positions
projected onto plane XY shown at birth of BPMG age. The positions
and sizes of the LCC and UCL shells are also shown. The lines
between the shells schematically show the flow created by the
compression. The symbols representing the stars as in Fig.
1.\label{fig5}}
\end{figure}

\begin{figure}
\epsscale{.80} \plotone{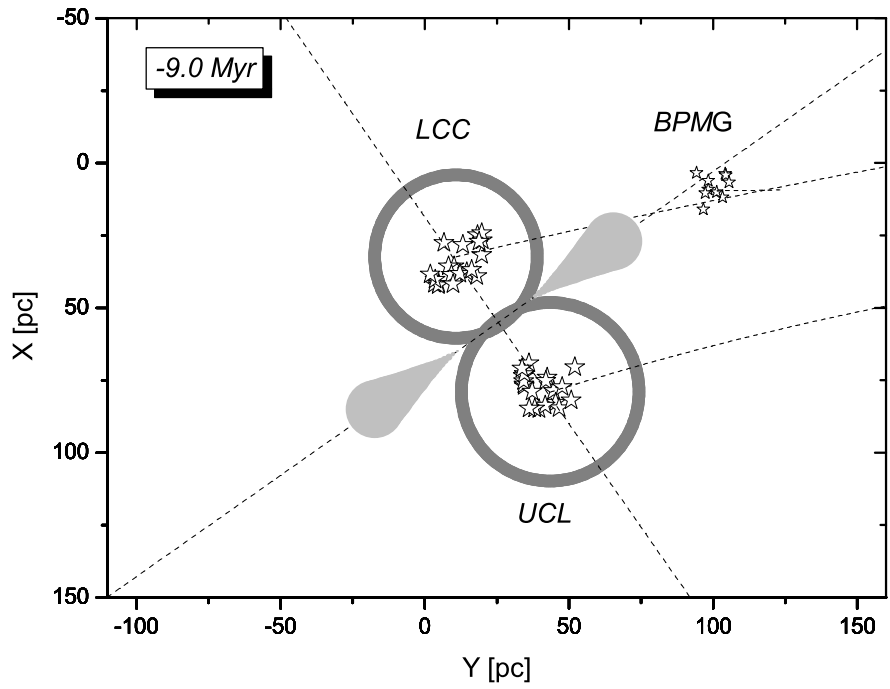} \caption{The same as in Fig. 5 but
at the time of the shells collision. The overlapping of the shells
is a projection effect. The symbols representing the stars as in
Fig. 1.\label{fig6}}
\end{figure}

\begin{figure}
\epsscale{.80} \plotone{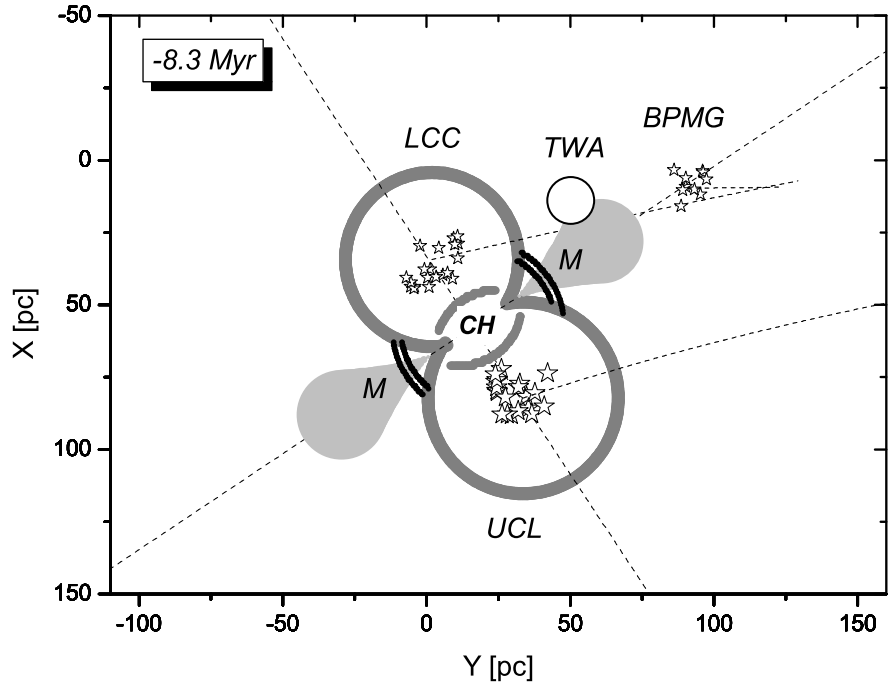} \caption{Positions of LCC and UCL
projected onto plane XY. The shells of these subcomponents are shown
at the age of 8.3 Myr when the TWA was born. The arcs inside the
bubbles schematically show the formation of the "champagne flows"
marked by symbol CH in the collision region. The Mach shocks (M) are
also schematically shown. The symbols representing the stars are as
in Fig. 1. \label{fig7}}
\end{figure}

\begin{figure}
\epsscale{.80} \plotone{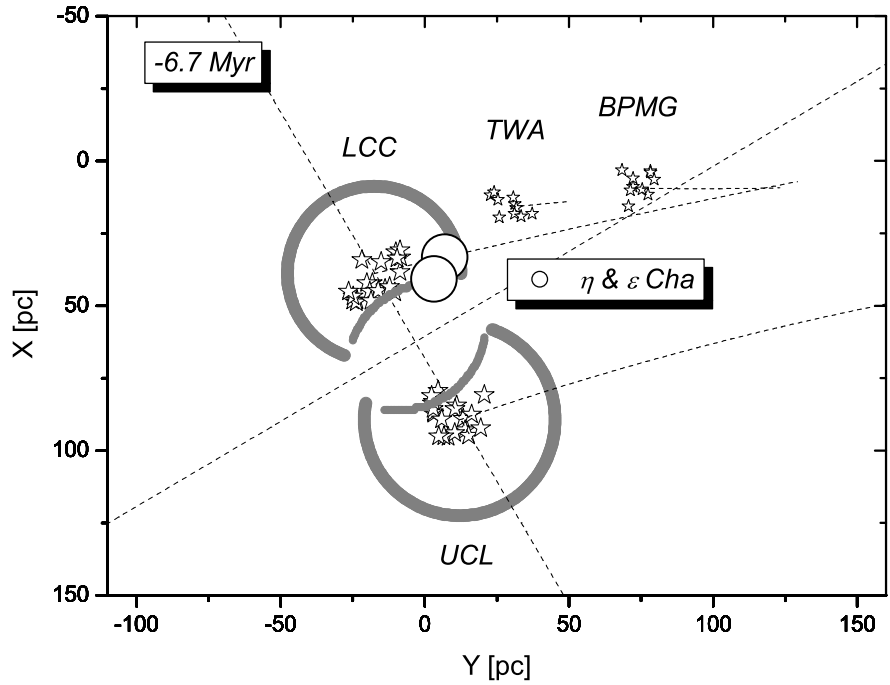} \caption{As in Fig. 7 but at -6.7
Myr. The place of birth of the $\eta$ Cha cluster at this age is
shown. The "champagne flows" propagate further into the bubbles with
a velocity that depends on the density of the medium. This is
schematically shown in the figure. The symbols representing the
stars as in Fig. 1. \label{fig8}}
\end{figure}

\begin{figure}
\epsscale{.80} \plotone{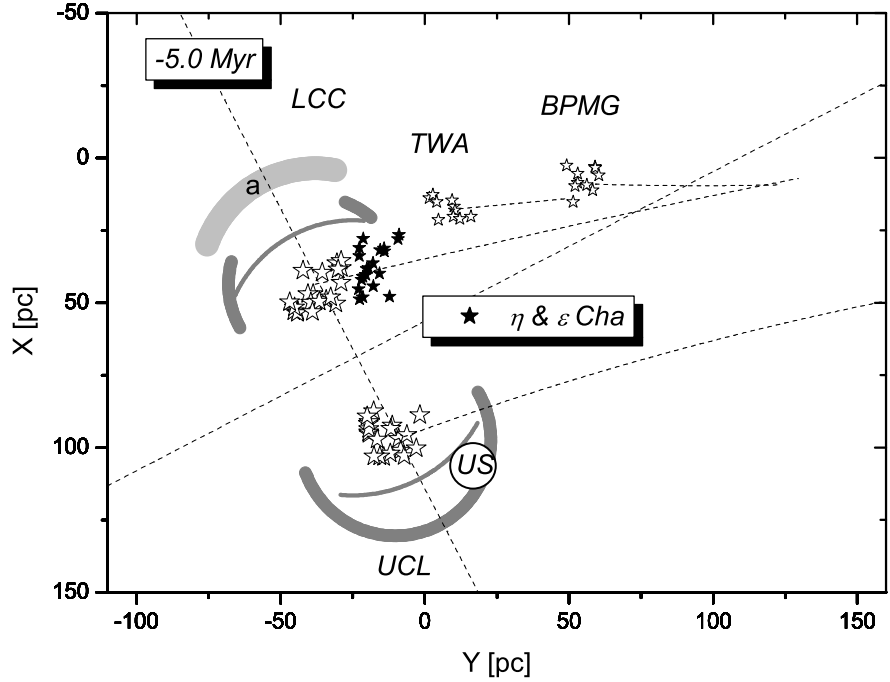} \caption{As in Fig. 8. but at -5.0
Myr, the age of US subcomponent. The place of this subcomponent is
shown. This location was determined on the basis of its 3D orbit. As
in Fig. 6 the projection on the plane XY shifts the 3D position of
US slightly into the bubble. The symbols representing the stars as
in Fig. 1.\label{fig9}}
\end{figure}

\begin{figure}
\epsscale{1.2} \plottwo {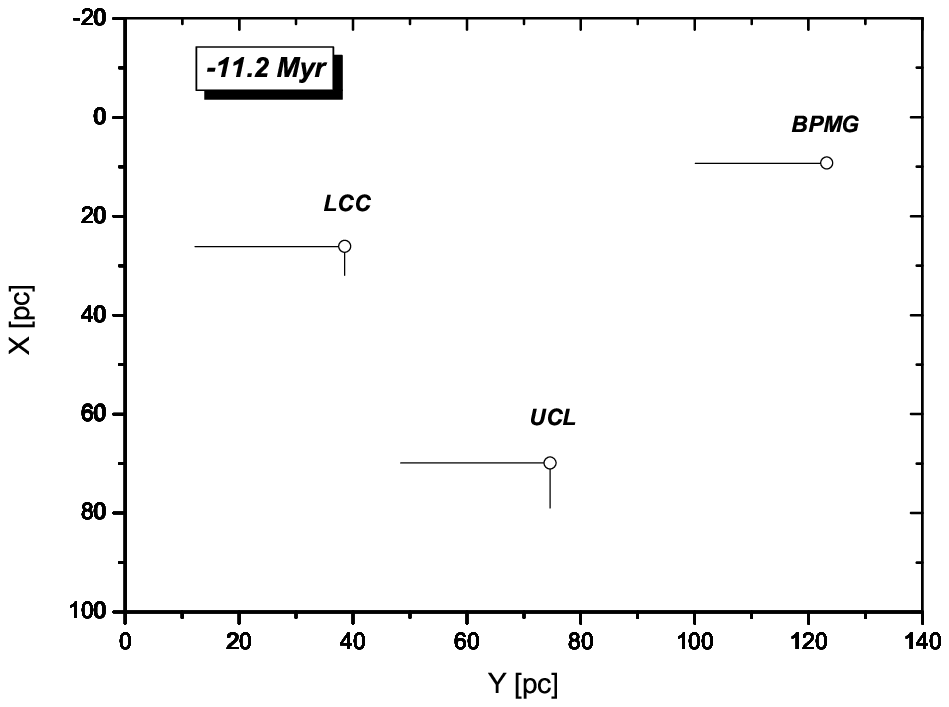}{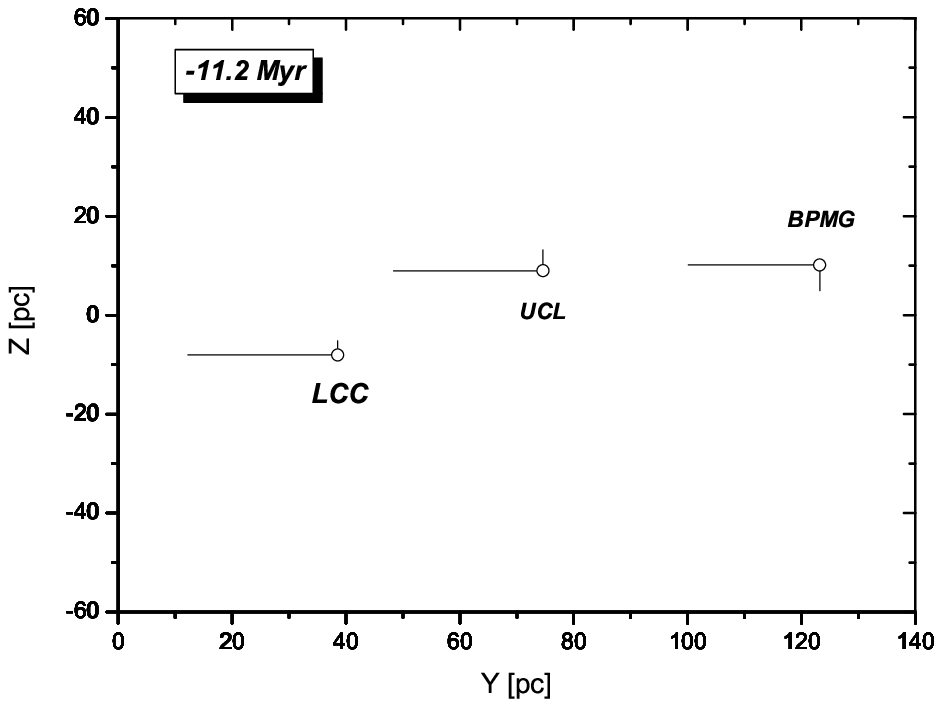} \caption{LSR positions
and space velocity components of BPMG at -11.2 Myr. The velocity
components are shown corresponding to a displacement along the orbit
in a time interval of 2 Myr.\label{fig6}}
\end{figure}

\begin{figure}
\epsscale{1.2} \plottwo {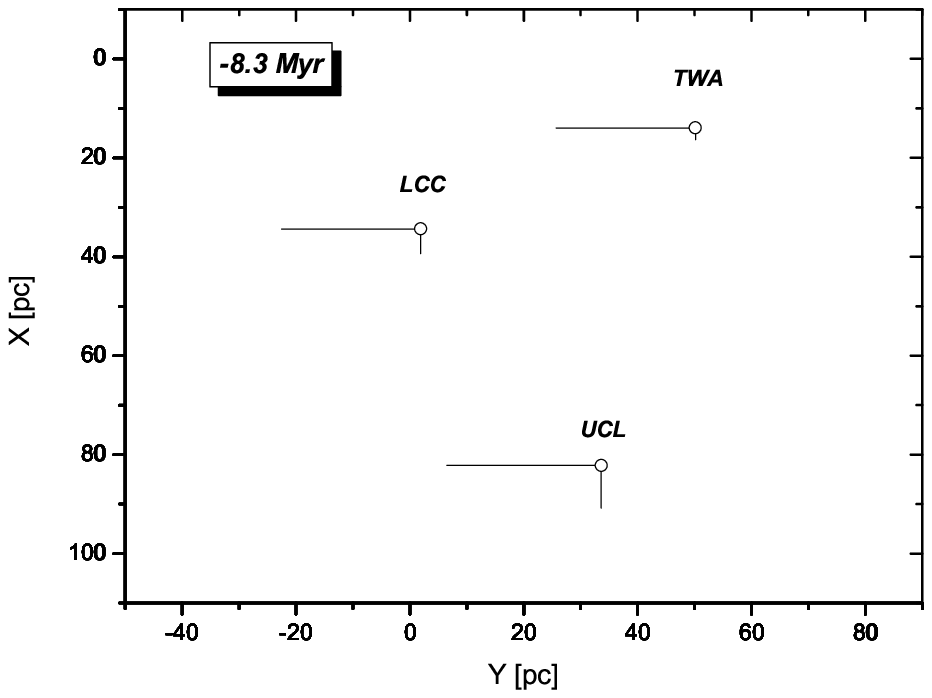}{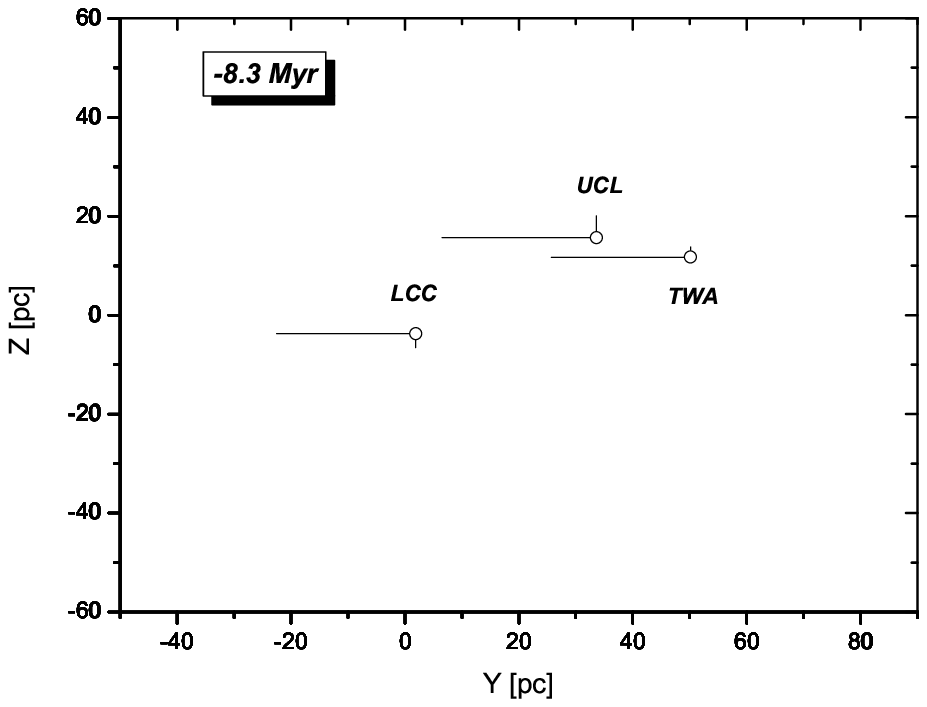} \caption{As at Figure
10 but for TWA. \label{fig6}}
\end{figure}

\begin{figure}
\epsscale{1.2} \plottwo {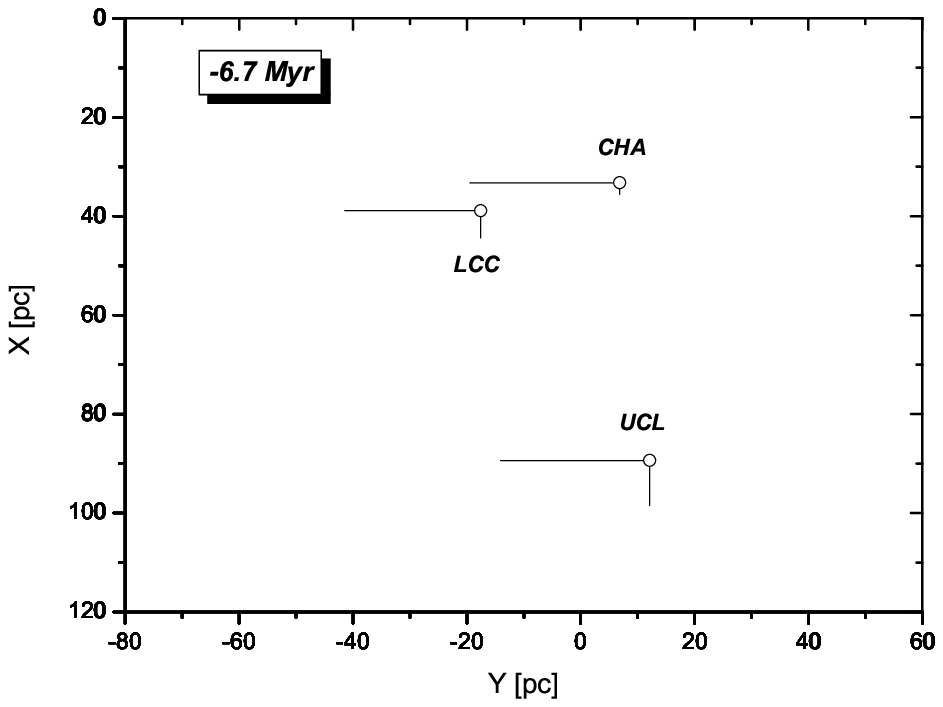}{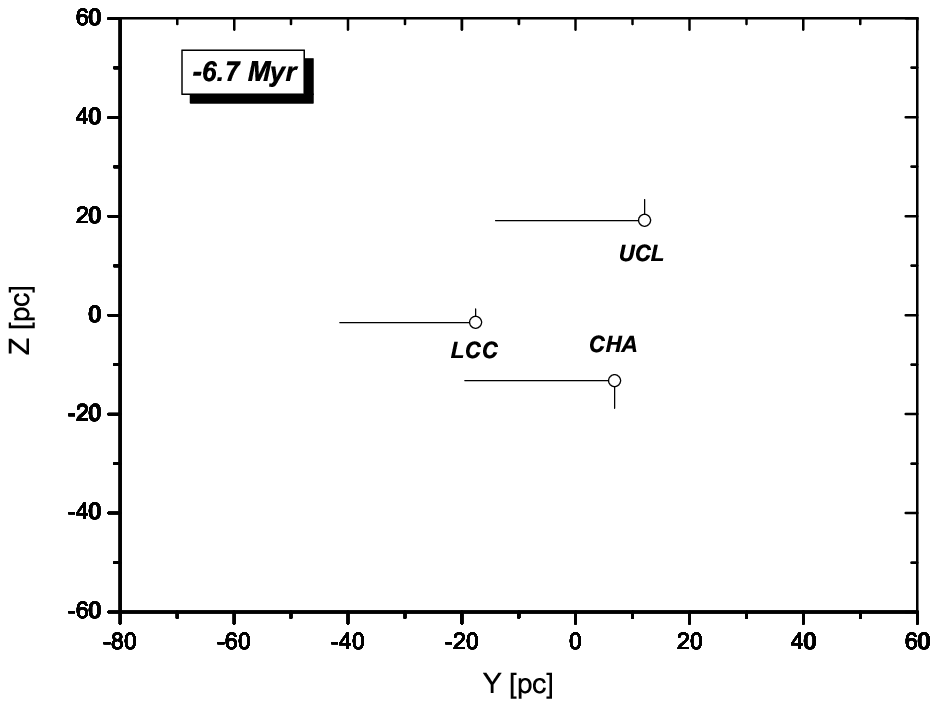} \caption{As at Figure
10 but for the $\eta$ Cha. \label{fig6}}
\end{figure}

\begin{figure}
\epsscale{1.0} \plottwo {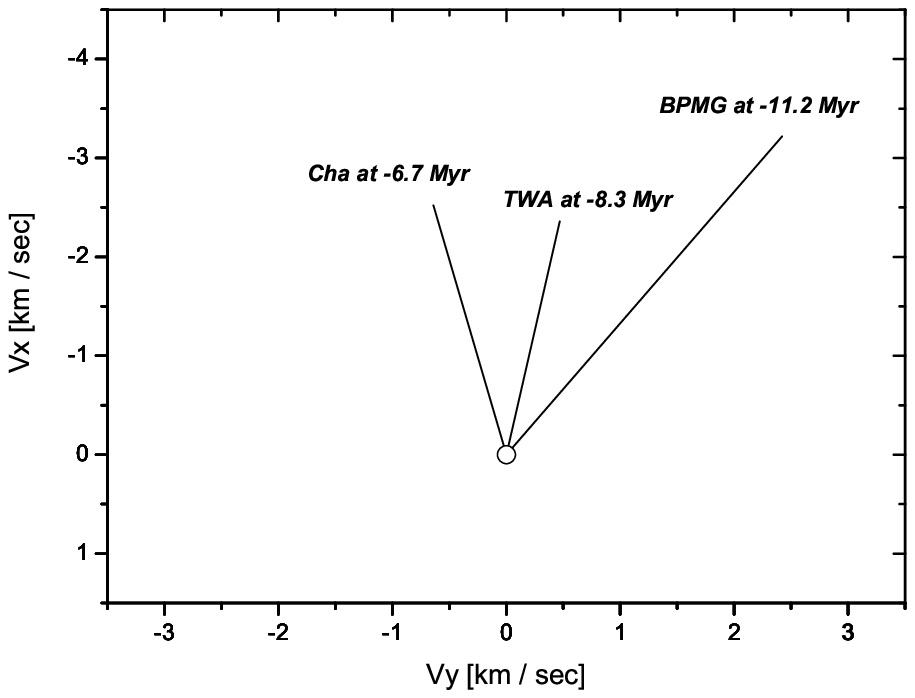}{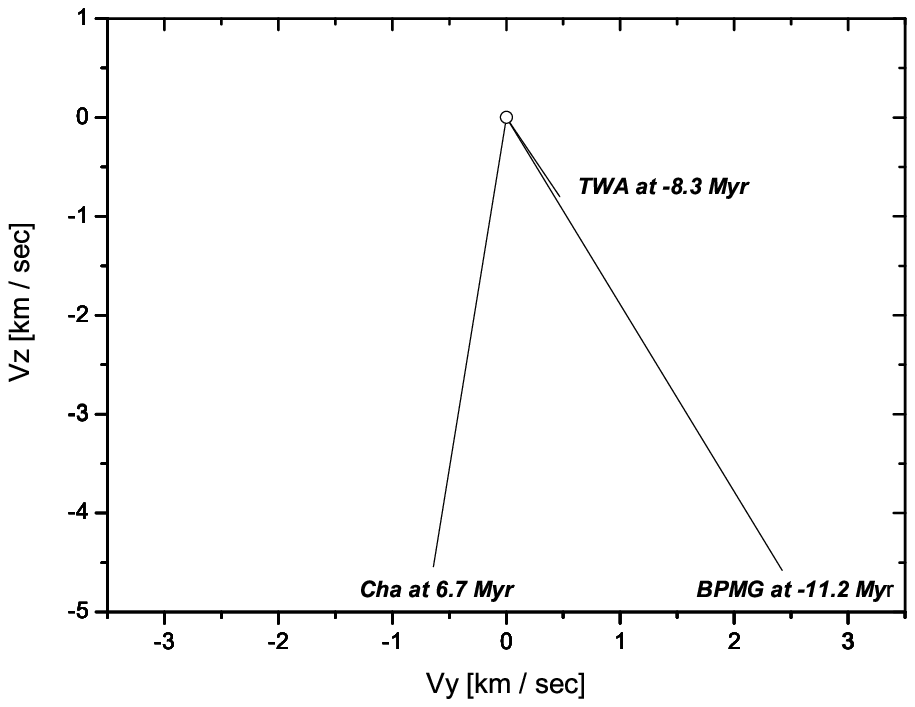} \caption{Space velocity
components of BPMG, TWA and $\eta$ Cha cluster relative to the
average motion of LCC and UCL at the epochs of their formation:
-11.2 Myr, -8.3 Myr and -6.7 Myr respectively. \label{fig6}}
\end{figure}

\begin{figure}
\epsscale{.80} \plotone{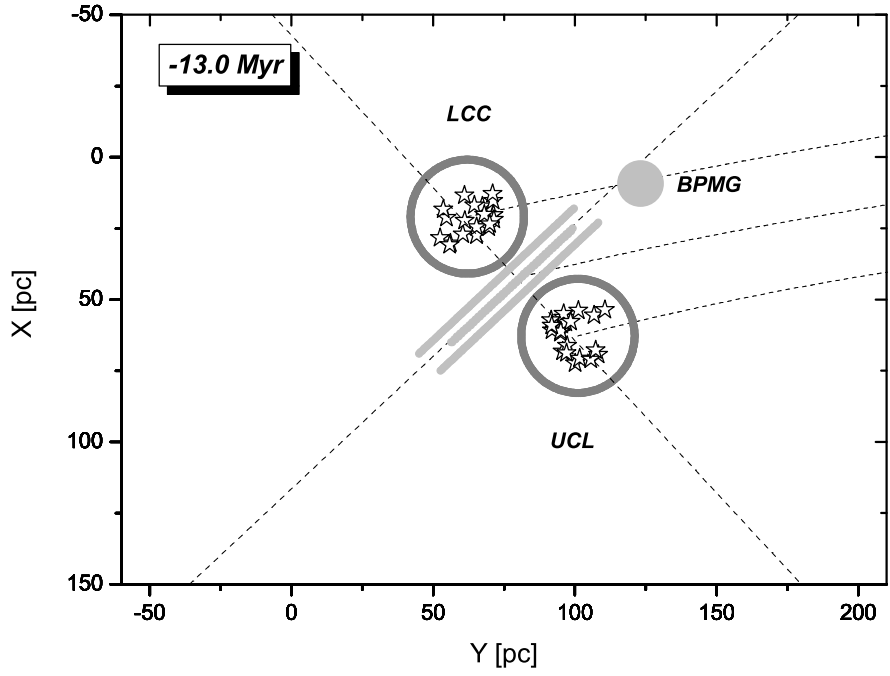} \caption{Space situation at the
epoch 13 Myr ago, when the common shells were formed and the region
where BPMG will be formed at about 11.2 Myr ago.\label{fig10}}
\end{figure}

\begin{figure}
\epsscale{.80} \plotone{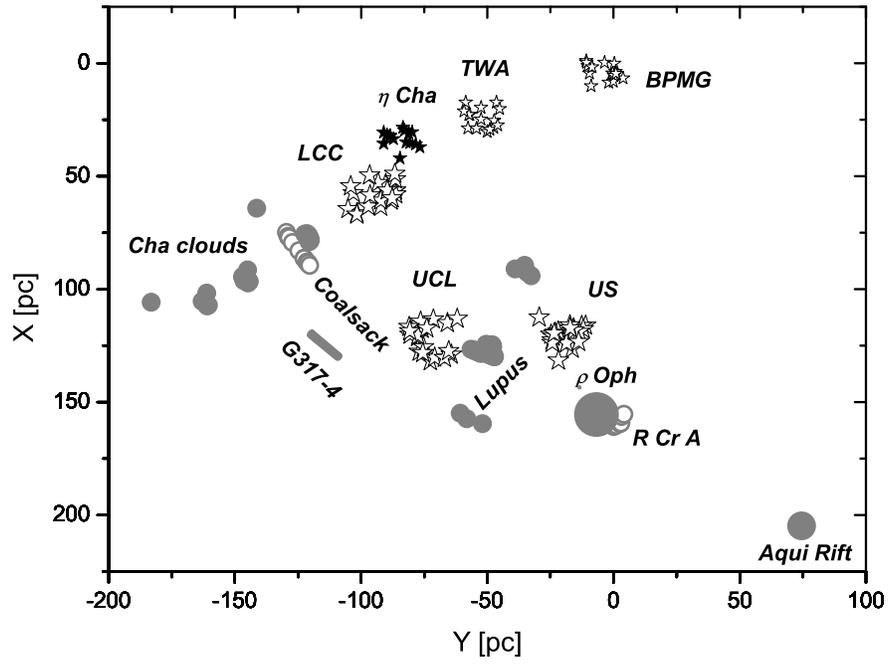} \caption{Present spatial positions
of the studied associations and moving groups.The grey symbols show
positions of CO molecular clouds in the Sco-Cen region (Dutra and
Bica 2002). The symbols representing the stars as in Fig.
1.\label{fig10}}
\end{figure}

\end{document}